\documentclass[aps,prx,twocolumn,notitlepage,superscriptaddress]{revtex4-2}
\usepackage{bm} \usepackage{graphicx} \usepackage{amsmath,amsthm,stmaryrd,amssymb} 
\usepackage{color}
\usepackage{hyperref}

\DeclareMathOperator*{\Tr}{Tr}

\newcommand{\ket}[1]{\vert{#1}\rangle}
\newcommand{\bra}[1]{\langle{#1}\vert}

\newtheorem{thm}{Theorem}

\makeatletter
\newcommand{\hathat}[1]{%
\begingroup%
  \let\macc@kerna\z@%
  \let\macc@kernb\z@%
  \let\macc@nucleus\@empty%
  \hat{\raisebox{.2ex}{\vphantom{\ensuremath{#1}}}\smash{\hat{#1}}}%
\endgroup%
}
\makeatother

\begin{document}

\title{Few measurement shots challenge generalization in learning to classify entanglement}

\author{Leonardo Banchi} \email{leonardo.banchi@unifi.it}
\affiliation{Department of Physics and Astronomy, University of Florence,
via G. Sansone 1, I-50019 Sesto Fiorentino (FI), Italy}
\affiliation{ INFN Sezione di Firenze, via G. Sansone 1, I-50019, Sesto Fiorentino (FI), Italy }
\author{Jason Pereira} 
\affiliation{ INFN Sezione di Firenze, via G. Sansone 1, I-50019, Sesto Fiorentino (FI), Italy }
\author{Marco Zamboni} 
\affiliation{Department of Physics and Astronomy, University of Florence,
via G. Sansone 1, I-50019 Sesto Fiorentino (FI), Italy}

\begin{abstract}
	The ability to extract general laws from a few known examples depends on the complexity 
	of the problem and on the amount of training data. In the quantum setting, the learner's  generalization 
	performance is further challenged by the destructive nature of quantum
	measurements that, together with the no-cloning theorem,
	limits the amount of information that can be extracted from each training sample. 
	In this paper we focus on hybrid quantum learning techniques where classical machine-learning methods 
	are paired with quantum algorithms and show that, 
	in some settings, the uncertainty coming from a few measurement 
	shots can be the dominant source of errors. 
	We identify an instance of this possibly general issue by focusing on the
	classification of maximally entangled vs.~separable states, showing that this toy problem 
	becomes challenging for learners unaware of entanglement theory. 
    Finally, we introduce an estimator based on classical shadows that performs better 
		in the big data, few copy regime.
	Our results show that the naive application of classical machine-learning methods to the quantum setting is 
	problematic, and that a better theoretical foundation of quantum learning is required.
\end{abstract}

\maketitle

{\it Introduction}:-- 
The ultimate performance of learning algorithms in the quantum setting is not 
completely understood in the general case. 
On one hand, classical learning bounds \cite{bartlett2021deep} predict that the generalization 
error, namely the excess error when trying to classify a new datum, unseen during training, 
should decrease as a function of the number $N$ of training states, typically
as $\mathcal O(N^{-1/2})$. Those bounds assume complete classical knowledge of
the training states and cannot be directly applied to quantum data. 
On the other hand, the ability to extract certain features from quantum data is limited by the number of 
available copies $S$ of those data. When the functional dependence of the data on the features is known
a priory, as in quantum estimation 
\cite{giovannetti2011advances,demkowicz2020multi}, the error 
typically scales as $\mathcal O(S^{-1/2})$ or even $\mathcal O(S^{-1})$ with some quantum strategies. 
Similarly, in quantum hypothesis testing the probability of error decreases exponentially with $S$,
and global strategies allows to achieve better exponents \cite{audenaert2008asymptotic}. 

In the more general setting where the functional dependence between the quantum data
and the features that we want to extract in unknown, we may ask ourselves what is 
the ultimate performance that we can achieve given  access to some known
examples, namely how does the error in classifying new
data behaves as a function of $N$, number of data, and $S$, number of copies, which 
can be linked to the number of measurement shots in local strategies. 
Although some partial results are available 
\cite{banchi2021generalization,banchi2024statistical,fanizza2024learning,caro2023out,zhao2023learning,monras2017inductive,guctua2010quantum,gil2024understanding,caro2022generalization,caro2023out,huang2020power,du2023problem,kubler2021inductive,cesa2011efficient,recio2024single,du2024efficient},
the general theory is still missing. For instance, 
Ref.~\cite{banchi2024statistical} found an upper bound 
on the error as $\mathcal O((NS)^{-1/3})$ using global strategies, which is independent on
of the dimension $d$ of the Hilbert space, and $\mathcal O(d (NS)^{-1/2})$ with local strategies based 
on partial tomography. 
As for local strategies, most of the quantum machine learning literature adapts 
techniques developed in classical settings by replacing some classical 
quantities with quantum expectation values \cite{schuld2021machine}.
How many measurement shots $S$ are needed to estimate those expectation values with 
the desirable precision in order to reproduce the expected classical behaviour? 
This problem is implicitly discussed in the barren plateau literature 
\cite{cerezo2021cost,thanasilp2022exponential}, since exponentially small expectation values 
require exponentially many shots. 
However, in learning scenarios, errors in the data are more tolerated, 
provided we have access to vast amount of training resources, so we might wonder whether we can reduce the 
error due to few measurement shots by simply increasing the dataset.

The lack of a complete understanding of how the errors in learning with quantum data behave 
as a function of $N$ and $S$ is also due to the difficulty of formalizing the problem 
when dealing with real-world datasets. 
Motivated by this, in this paper we consider a simple binary classification task
of deciding whether a pure state is separable or maximally entangled. 
In spite of the simplicity of the problem, which also admits an analytic solution, 
we show that learning algorithms are very inefficient, both in terms of the scaling with 
$N$ and $S$. 
We focus on two related algorithms: one uses classical recipes (support vector 
machines \cite{chang2011libsvm}) by focusing on the individual states in the training set,
while another one follows the quantum hypothesis testing idea of dealing with uncertainties 
by introducing suitable mixed states. 
We show that both algorithms fail to learn to classify entanglement directly from data, 
in spite of the simplicity of the problem.

Our analysis defines a simple example to study the performance of learning algorithms 
as a function of the available resources, e.g.,~data and shots, 
which is easy to interpret and yet challenging from a learning perspective.

{\it Problem definition}:-- 
We focus on the toy problem of classifying whether a pure quantum 
state belongs to either of two classes, 
the class of separable states or the class of maximally entangled states. 
These two classes are generated by 
applying arbitrary local operations on two different reference states 
of a bipartite system
\begin{equation}
\begin{aligned}
	\ket{\psi^{\rm sep}} &= U_A\ket0_A\otimes U_B\ket{0}_B,\\
	\ket{\psi^{\rm ent}} &= U_A\otimes U_B \ket{\Phi}_{AB}.
\end{aligned}
\label{eq:states}
\end{equation}
where the two subsystems have indices $A$ and $B$. 
In the above definition, 
$\ket{\Phi}_{AB} = d^{-1/2}\sum_{i=1}^d\ket{i,i}_{AB}$ is 
the reference maximally entangled state, and 
$U_{A/B}$ are possibly different local unitaries.

We now introduce the following learning problem. A learner is given 
an unknown state 
\begin{equation}
\rho^y = \ket{\psi^y}\!\bra{\psi^y},
\end{equation}
with the premise that it belongs either to the class of separable 
states, for which we assign $y=1$, or to the class maximally 
entangled states, for which we assign $y=-1$, and the learner's task 
is to extract the unknown 
label $y$.

If the learner does know the theory of entanglement, then this is 
a trivial problem. 
Indeed, it is always possible to measure the purity of either subsystem 
$A$ or $B$, or their R\'enyi entropy, to assess whether the state was 
entangled or not \cite{horodecki2009quantum,harrow2013testing}. 
The purity $\Tr[(\rho_A^y)^2]$ is equal to the value obtained by 
measuring the SWAP observable
\begin{equation}
	\Tr\left[(\rho_A^y)^{\otimes 2}{\rm SWAP}\right] =
	\begin{cases}
		1 & {\rm ~for~} y=+1,\\
		\frac1d & {\rm ~for~} y=-1,\\
	\end{cases}
	\label{eq:swap}
\end{equation}
where $\rho^y_A=\Tr_B[\rho^y]$ and the same result applies for 
the partial trace on the other subsystem. 
For separable states the SWAP measurement will have zero variance, 
while for entangled states, repeating the measurement $S$ times 
the variance goes as $(1-d^{-2})/S$. Therefore, few measurement 
shots $S$ are required to assess with almost certainty whether 
the state was entangled or not, and
the problem gets easier for larger $d$, since the separation between 
the two outcomes in Eq.~\eqref{eq:swap} increases.

{\it Learning formulation}:-- 
We now show that the above simple problem becomes challenging for a 
quantum learner that does not know the theory of entanglement, nor 
the symmetries of the problem, and that
is is not aware of the existence of the simple discrimination 
operator shown in Eq.~\eqref{eq:swap}. 

We operate in a supervised learning framework and 
assume that the learner has access to a training set made of $2N$
pairs $(\rho^{y_n}_n,y_n)$, composed of either a separable or 
entangled state, and their label, equally split so that 
the number of states with $y=\pm1$ is the same. 
The states are unknown, meaning that 
they are stored into a quantum memory and the learner has no prior classical 
description of them. However, the learner can create multiple copies of each
training state, by repeating a known state preparation routine, 
and manipulate all of them, either individually or jointly
with a quantum algorithm. The
goal of the learner is to use the information contained in the training 
set to construct an observable, or an algorithm, capable of certifying whether a new 
state $\rho^y$, possibly not present in the training set,  is entangled or not. 
The construction of this observable is challenged not only by the finite 
number of training pairs $N$, but also by limited information that 
the learner can extract about the unknown training states with a finite 
number of copies of each state. 

To simplify the learner's task, we tell them to search for observables 
acting on two copies of the state, as in Eq.~\eqref{eq:swap}. 
More generally, the discrimination function could be non-linear in the state.
To simplify the function class, we restrict ourselves to polynomial functions of $\rho$ of order $c$; 
such functions, in the most general case, can be rephrased as a ``linear'' measurement over 
$c$ copies of the state. Therefore, we choose 
\begin{align}
	y&= {\rm sign}\left(f\left[(\rho^y)^{\otimes c}\right]\right), &f(\rho)&=\Tr\left[A\rho\right],
	\label{eq:yfx}
\end{align}
namely where the decision is based on the sign of the expectation 
value of an observable $A$, that we have to learn from the training data. 
Using Eq.~\eqref{eq:swap}, we know that for $c=2$ the model has an
exact solution
\begin{equation}
	A_* \propto \frac{d}{d-1}\left({\rm SWAP}_{A_1A_2} + {\rm SWAP}_{B_1B_2} \right) 
		- \frac{d+1}{d-1}\openone,
		\label{eq:representer}
\end{equation}
where the coefficients have been chosen to maximally separate the 
``margin'' and normalized to have $\Tr[A_*(\rho^y)^{\otimes 2}]=y$,
without the need of the sign function. 
${\rm SWAP}_{A_1 A_2}$ swaps the $A$ modes and acts as an identity on the $B$ modes, and vice versa for ${\rm SWAP}_{B_1 B_2}$. Both terms are included so that the observable is symmetric under swapping the $A$ and $B$ systems.

Since the learner is unaware of an exact solution such as \eqref{eq:representer},
during {\it training}
they try to approximate $A_*$ by minimizing 
a regularized empirical loss function
\begin{equation}
	L_{\rm emp} = \frac1{2N}\sum_{n=1}^N \sum_{y_n=\pm1}\ell(y_n,f\left[(\rho_n^{y_n})^{\otimes c}\right]) + g(\|A\|).
	\label{eq:L emp}
\end{equation}
The first term in $L_{\rm emp}$ is the average loss function 
over the $2N$ elements of the training set, where the loss 
$\ell$ quantifies the discrepancy 
between the predicted $y$ from Eq.~\eqref{eq:yfx} and the real label $y_n$, 
which is known since $\rho_n$ is in the training set. 
The second term introduces a penalty $g(\|A\|)$, a function $g$ of the 
norm of the observable $A$, which helps constrain the set of observables 
to avoid overfitting \cite{banchi2024statistical}.

A popular learning model called a support vector machine (SVM) uses 
the hinge loss function and an $L_2$ penalty 
\cite{chang2011libsvm,schuld2021machine,liu2021rigorous}. There it is possible to prove that 
the problem admits 
a dual formulation where the solution can be expressed as a linear 
combination of the training data,
\begin{equation}
	A_*^{\rm emp} = \sum_{n=1}^N \sum_{y_n=\pm1}\alpha_n y_n (\rho^{y_n}_n)^{\otimes c} + \beta \openone,
	\label{eq:decision obs emp}
\end{equation}
with some coefficients $\alpha_n$ and $\beta$. This explains our choice of the function class in Eqs.~\eqref{eq:yfx}, 
which, due to linearity, may allow us to reach \eqref{eq:representer} with suitable coefficients when $N\to\infty$. 
For any finite $N$, the coefficients $\alpha_n$ can be numerically
obtained using open-source libraries \cite{chang2011libsvm} by solving the 
following convex optimization problem
\begin{equation}
	\max_{\{0\leq \alpha_n\leq C\}} \sum_n \alpha_n - \!\!\sum_{n,m,y_n,y_m}\!\! \frac{
	\alpha_n\alpha_my_ny_m}2 K_c(\rho_n^{y_n},\rho_m^{y_n}),
	\label{eq:svmtraining}
\end{equation}
for a given constant $C$ (we use the default value $C=1$), while 
$\beta$ can be expressed in terms of the learnt $\alpha_n$. 
In the above equation we have defined the ``kernel'' 
for an arbitrary number of copies $c$ as
\begin{equation}
	K_c(\rho,\rho') = \Tr[\rho^{\otimes c}\rho'^{\otimes c}] = 
		|\bra{\psi}\psi'\rangle|^{2c},
		\label{eq:kernel}
\end{equation}
and the irrelevant dependence on $y$ has been dropped to simplify the
notation.

Inserting Eq.~\eqref{eq:decision obs emp} in \eqref{eq:yfx} we can 
rewrite the classifier of a new ``test'' state $\rho$ as a
{\it kernel expansion of the data} \cite{schuld2021machine}
\begin{equation}
	f(\rho^{\otimes c}) = \sum_{n=1}^N \sum_{y_n=\pm1}\alpha_n y_n K_c(\rho_n^{y_n},\rho)+\beta.
	\label{eq:svmtesting}
\end{equation}
Both training,  via Eq.~\eqref{eq:svmtraining}, and testing, 
via Eq.~\eqref{eq:svmtesting}, require the estimation of the kernel entries in Eq.~\eqref{eq:kernel}. 
This can be done in multiple ways, using swap tests \cite{schuld2021machine} with $S$ measurement shots,  
global strategies involving $S$ copies \cite{fanizza2020beyond} or via classical shadow techniques \cite{anshu2022distributed}.
We focus on the swap test, due to its simplicity. Per Eq.~\eqref{eq:kernel}, estimating $K_c(\rho,\rho')$ is 
equivalent to estimating $K_1(\rho,\rho')$ (the overlap of the states) and taking the $c^{\rm th}$ power as a post-processing step. 
This is allowed due to the tensor product form of the states; if either state were not in this form, we would 
need to perform a $c$-system swap test.

\begin{figure}[t]
	\centering
	\includegraphics[width=0.45\textwidth]{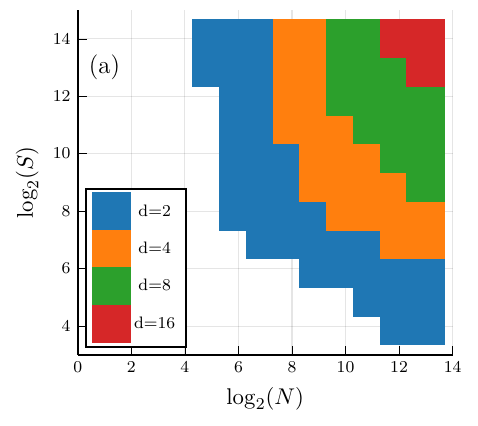}
	\includegraphics[width=0.45\textwidth]{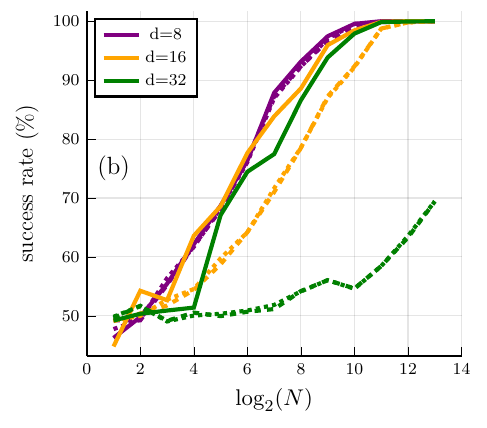}
	\caption{
		(a) Region with success rate higher than 99\% in classifying a new state, not present in the training set, 
		vs.~number of training pairs $N$ and number of shots $S$, for different dimensions $d=2,4,8,16$. 
		(b) Comparison between success rate in classifying a new state and $N$, for $d=8,16,32$. 
		Solid lines show numerical 
		simulations with exactly computed expectation values ($S\to\infty$), dashed lines use $S=2^{14}=16384$ shots, 
		while dotted lines (mostly overlapping with the dashed ones) also use $S=2^{14}$ shots, but then directly construct the ``analytical'' classifier ($B_{\mathrm{obs}}$, the unbiased estimator of $B^{}_{(2)}$ from Eq.~(\ref{eq:opt B})) from the measurement results, rather than finding it using a support vector machine.
	}
	\label{fig:success}
\end{figure}

{\it Numerical results:--} Numerical results are shown in Fig.~\ref{fig:success}, 
where the accuracy of the decision observable \eqref{eq:decision obs emp},
which results in decision function \eqref{eq:svmtesting}, is 
tested over new states. The full numerical results are presented in Appendix~\ref{sec:more}. 
Both the test states and the $2N$ training states are randomly generated following 
Eq.~\eqref{eq:states}, 
with Haar-random choices of the matrices $U_A$ and $U_B$. 
Each kernel entry for training,  Eq.~\eqref{eq:svmtraining}, and testing, 
Eq.~\eqref{eq:svmtesting}, is obtained by estimating $K_1$ 
via a SWAP test \cite{schuld2021machine} with $S$ measurement 
shots, and taking the $c$th power, 
though better precision can be obtained with global strategies \cite{fanizza2020beyond}. 
Here we consider $c=2$ and different dimensions $d$ 
of the Hilbert spaces of the two subsystems in \eqref{eq:states}.
In Fig.~\ref{fig:success}(a) we plot the values of $S$ and $N$ for which the learnt classifier 
is able to classify new states with accuracy higher than 99\%. We see that the high-accuracy 
region shrinks for larger $d$, to the point of being empty for $d=32$, in spite of 
the large number of training data and large number of shots (both about $\sim16000$). 
The difference with respect to the classical literature is further emphasized in 
Fig.~\ref{fig:success}(b), where we compare the performance 
of the classifier with the largest number of shots ($2^{14}$) with the noiseless one. 
We see that for $d=8$ the performance of the two is comparable, they start to differ for 
$d=16$ and are very different for $d=32$. 
In summary, the poor performance of the $d=32$ case is not due to the lack of data, but to the lack 
of sufficient measurement shots. 
This numerical analysis shows that, even for small dimensional systems, the learner is not 
able to find an observable capable of successfully discriminating separable and maximally 
entangled states, although such an observable exists and is rather simple -- see Eq.~\eqref{eq:representer}.

Therefore, in the ``big-data'' regime, where $N$ is very large, a limited number of measurement shots 
might be the dominant source of errors. 
In some sense, this is due to the fact that the errors in the kernel entries, due to a finite $S$,
are amplified in the solutions $\alpha_n$ of the optimization problem \eqref{eq:svmtraining}.
This shows that naively applying classical machine learning techniques to the quantum case, without taking into 
account the errors introduced by the quantum measurements, may result in large generalization errors. 

We can try to understand the dependence on $S$ from Fig.~\ref{fig:success} by noting 
that, before taking the $c$th power, on average, 
$K_1(\rho,\rho')=\mathcal O(d^{-2})$ while the expected 
error in the swap test with $S$ shots is $O(S^{-1/2})$. 
Therefore, we need at least 
$\log_2S\gg 4\log_2 d$ to estimate each kernel entry with enough precision.
As for the dependence on $N$, we can study the generalization error, namely 
the difference between the training loss and the average test loss, assuming full knowledge of the 
training states. 
Thanks to the result of \cite{banchi2021generalization,banchi2024statistical}, 
under the most general measurements, not restricted to the swap test,
this error is upper bounded by $\sqrt{\mathcal B/N}$ where $\mathcal B = \Tr[\sqrt{\bar \rho}]^2$
and $\bar\rho$ is the average state. The explicit calculation is performed 
in Appendix \ref{sec:c2}, where we find that $\mathcal B =\mathcal O(d^4)$. 
Therefore, in order to have a small generalization error we need a number of training data $N$ 
that satisfies $N\gg \mathcal O(d^4)$. 
This condition can also be understood by noting that the difference between the empirical loss 
over the training set and the average loss over the whole state distribution goes to zero 
if the training states form a $2$-design. Since $2$-designs require at least $\mathcal O(d^4)$ 
unitaries \cite{gross2007evenly}, the number of training states cannot be smaller.
In summary, by separately considering 
the errors due to the finite number of shots $S$ and due to the finite amount of data $N$, we need to enforce 
$ 
	\log_2(S) \gg \mathcal O(4 \log_2 d), ~\log_2(N) \gg \mathcal O(4 \log_2 d),
$ 
which seems too pessimistic, according to Fig.~\ref{fig:success}. Indeed, we expect that a larger 
$N$ might mitigate, to some extent, the measurement shot noise, but the above separate treatment does not 
allow for quantitative predictions. 

{\it Understanding errors:--}
To have a better understanding of the combined errors, we introduce a strategy inspired by 
quantum hypothesis testing, based on working directly with averages, rather than 
single training states. This is made possible by the linearity of the problem in the multi-copy state space.
Following \cite{banchi2024statistical},
by exploiting the convexity of the hinge loss $\ell$ we may bound \eqref{eq:L emp} as 
\begin{align}
	L_{\rm emp} 
 \leq \sum_y  \ell\left(y,f(\hat\rho^y_{(c)})\right) + g(\|A\|), 
 \label{eq:Loss emp ineq}
\end{align}
where 
\begin{equation}
	\hat \rho_{(c)}^{y} = \frac1N \sum_{n=1}^N (\rho_n^{y})^{\otimes c}. 
	\label{eq:emp state}
\end{equation}
In appendix~\ref{sec:error} we are then able to prove the following 

\begin{thm}\label{thm:optimal}
The optimal observable, which minimizes the upper bound \eqref{eq:Loss emp ineq} with an $L_2$ penalty term,
converges to Eq.~\eqref{eq:representer} for $N\to\infty$ and $c=2$. 
\end{thm}
Since support vector machines optimize the lower bound in \eqref{eq:Loss emp ineq}, 
the above theorem also implies that their solution will converge to an optimal classifier 
in the limit of infinitely many data and 
perfect kernel estimates. 

In the limit of infinite data and copies, the optimal classifier 
from Theorem~\ref{thm:optimal} can also be 
obtained by measuring the following observable 
\begin{equation}
	B^{}_{(2)}= 2\left(\bar \rho^{+}_{(2)}-\bar\rho^{-}_{(2)}\right)-\left(\bar \Delta^{}_{_{++}}-\bar \Delta^{}_{_{--}}\right)\openone,
	\label{eq:opt B}
\end{equation}
where the ``bar'' symbol denotes abstract averages in the 
limit $N,S\to \infty$. Classification is then carried out according to the sign of the measurement result. When $N$ and $S$ are finite, 
we can define an estimator $B_{\mathrm{obs}}$ of 
$ \Tr\big[\rho^{\otimes 2} B^{}_{(2)}\big]$
as a two step process -- see Appendix~\ref{sec:single copy}.
The first step, ``training'', consists in estimating the 
kernel entries $\bar \Delta^{}_{{yy'}}=\Tr[\bar \rho^{y}_{(2)}\bar \rho^{y'}_{(2)}]$ 
between average training states with labels $y$ and $y'$,
using $\mathcal O(SN^2)$ swap tests.  
The second step, ``testing'' of a new state $\rho^y$, consists in using another
$\mathcal O(NS)$ swap tests between $\rho^y$ and each training state. 
Each swap test between two states $\rho$ and $\rho'$ uses a single copy 
of each state. 
We then prove 
\begin{thm}\label{thm:bias var}
$B_{\mathrm{obs}}$ is an unbiased estimator of 
$ \Tr\big[\rho^{\otimes 2} B^{}_{(2)}\big]$
with expected variance over test states 
\begin{align}
	\mathbb{E}_{\mathrm{test}}\left[\mathrm{Var}_{N,S}[B_{\mathrm{obs}}]\right] &= 
			\mathcal O\left(\frac1N\left(\frac1S+\frac{1}{d^4}\right)^2\right)
			:= \sigma^2. 
\end{align}
\end{thm}
\noindent 
We define $\mu = |\mathbb{E}_{\mathrm{in-class~test},N,S}[B_{\mathrm{obs}}]|$, where the expectation value is taken over test states from the same class (and all choices of $N$ training states and $S$ measurement shot results).
Since $\mu= 
\mathcal O\left(d^{-4}\right), $ the error in the estimation must 
be much smaller than this. From Cantelli's inequality, we can bound 
the probability of misclassification as $\epsilon \simeq (\sigma/\mu)^2$. 
Then the above theorem implies that 
accurate simulation is possible for large $d$ provided that 
\begin{equation}
	N = \frac1{\epsilon} \mathcal O \left(1+\frac{d^4}{S}\right)^2,
\end{equation}
In the limit $S\to\infty$, or more precisely when $S=\mathcal O(d^4)$, 
$N=\mathcal O(\epsilon^{-1})$ data are sufficient, 
irrespective of the dimension $d$, a prediction compatible with Fig.~\ref{fig:success}(b). 
On the other hand, when $S\ll d^4$ we get 
$ 
	NS^2 \simeq d^8 \epsilon^{-1}.  
$ 
This shows that, for large $d$ and finite $N$ and $S$, the error 
arising due to the finite number of measurement shots may be 
the dominant source of generalization error. 
For $N\gg \mathcal O(d^8) $, the error goes to zero irrespective of $S$. 
Intuitively, this comes from the fact that the estimator $B_{\mathrm{obs}}$ 
requires $\mathcal O(N^2S)$ swap tests, 
so even if the single swap test has a large measurement noise, due to a small $S$, 
these errors are averaged out and do not affect much the learnt coefficients, 
unlike those in the solution of \eqref{eq:svmtraining}. 

The performance $B_{\mathrm{obs}}$ from Theorem~\ref{thm:bias var} 
is tested in Fig.~\ref{fig:success}(b), 
where we observe almost identical results to those obtained with support 
vector machines. However, such an estimator is computationally much faster for large $N$, 
since it does not require solving Eq.~\eqref{eq:svmtraining}.

{\it Learning with shadow measurements:--}
The above algorithm uses $\mathcal O(NS)$ copies of each training state 
during training and $\mathcal O(S)$ copies of each training state during testing, in order to estimate the kernels via 
the swap test. An alternative route \cite{anshu2022distributed} 
involves the use of randomized 
measurements to extract a classical representation of each state - 
its classical shadow - that can be classically manipulated during 
both training and testing. The advantage of this approach is that no copies 
of the training states are required for testing new data, and that each state 
can be measured independently, without having to load two different states in a quantum register. 
The algorithm starts by sampling $N_U$ unitaries $U_i$ from the Haar distribution and,
for each $d^2$-dimensional state $\rho$, 
collects $N_M$ measurement outcomes $x_j \sim \bra{x}U_i^\dagger\rho U_i \ket{x}$, 
with $x_j=0,\dots,d^2-1$ and $j=1,\dots,N_M$. The $N_M N_U$ outcomes define the classical shadow of the 
state $\rho$, whose construction requires $S = N_M N_U$ copies of $\rho$. 
From the classical shadows of two $d^2$-dimensional states 
$\rho_1,\rho_2$, like the ones from Eq.~\eqref{eq:states}, 
the algorithm from \cite{anshu2022distributed} 
estimates the overlap $\Tr[\rho_1\rho_2]$ with variance $\sigma^2$, provided that
$N_U\geq \max\{1,(d\sigma)^{-2}\}$ and $N_M \geq N_U^{-1/2}d/\sigma$.
Theoretical analysis of the error scaling for this algorithm is left to future studies, however
numerical results are presented in Ref.~\cite{Zamboni} and Appendix~\ref{sec:more}.

{\it Discussion:--}
We have shown that the naive application of tools from classical machine learning to the quantum case 
may lead to large generalization errors. This is due to the destructive nature of quantum measurements, 
which limit the amount of information that can be extracted from each training sample with a few measurement shots. 
We have identified a toy problem where this error is significant, namely the classification 
maximally entangled and separable states. 
We have shown 
that finding the optimal classifier  is difficult when the learner 
only has access to data (in the form of unknown quantum states), without any
prior knowledge of  entanglement theory. 
Moreover, in some cases the generalization error is dominated by the measurement shot error. 

Although we focused on a toy learning problem with quantum data, similar conclusions 
can be observed when using quantum kernel methods to classify classical data
\cite{thanasilp2022exponential,huang2020power}, without a careful choice of the 
feature map.

In general, our results show that standard strategies from classical machine learning, e.g.,~based on 
empirical risk minimization, might not be directly applicable to the quantum case without 
encountering a significant measurement overhead. Given also that quantum states cannot be copied, 
this requires rethinking the sample complexity bounds in the quantum case \cite{banchi2024statistical}.

{\it Acknowledgments:--}
The authors acknowledge financial support from: PNRR Ministero Università e Ricerca Project 
No. PE0000023-NQSTI funded by European Union-Next-Generation EU (L.B:); 
Prin 2022 - DD N.~104 del 2/2/2022, entitled ``understanding the LEarning process of QUantum
Neural networks (LeQun)'', proposal code 2022WHZ5XH, CUP B53D23009530006 (L.B.); 
U.S. Department of Energy, Office of Science, National Quantum Information Science Research
Centers, Superconducting Quantum Materials and Systems Center (SQMS) under the Contract
No. DE-AC02-07CH11359 (J.P. and L.B.).

\appendix

\begin{figure*}[t!]
	\centering
	\includegraphics[width=0.40\textwidth]{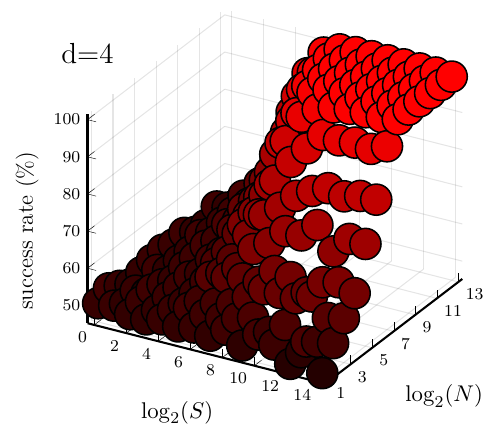}
	\includegraphics[width=0.40\textwidth]{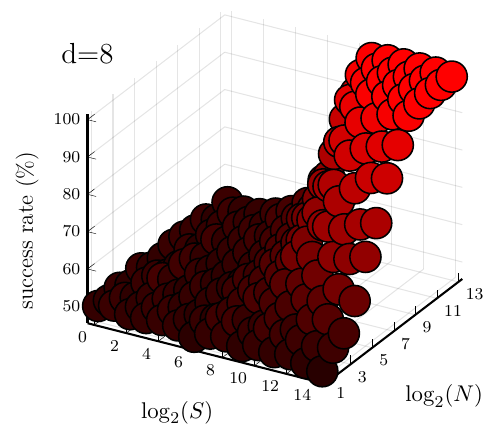}
	\includegraphics[width=0.40\textwidth]{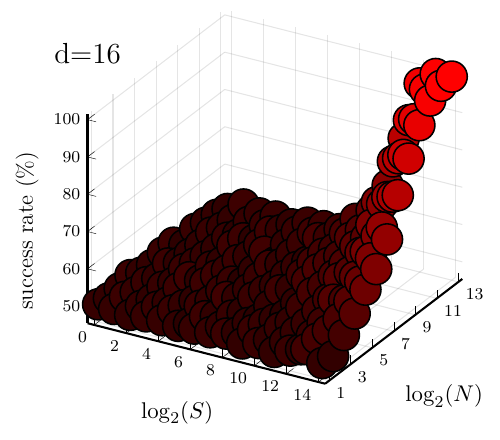}
	\includegraphics[width=0.40\textwidth]{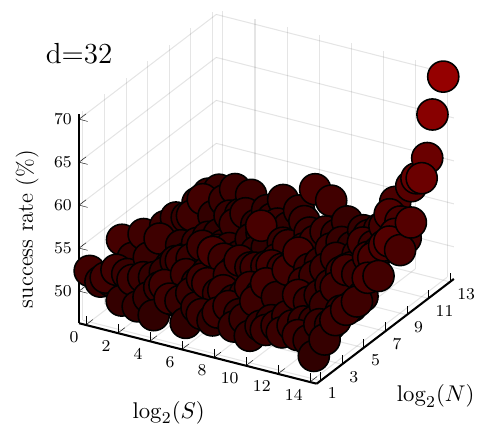}
	\caption{Success rate in learning to classify a new state, not present in
		the training set, vs number of shots $S$ and number of training pairs
		$N$ for different dimensions $d=4,8,16,32$.}
	\label{fig:result}
\end{figure*}

\begin{figure}[t]
	\centering
	\includegraphics[width=0.45\textwidth]{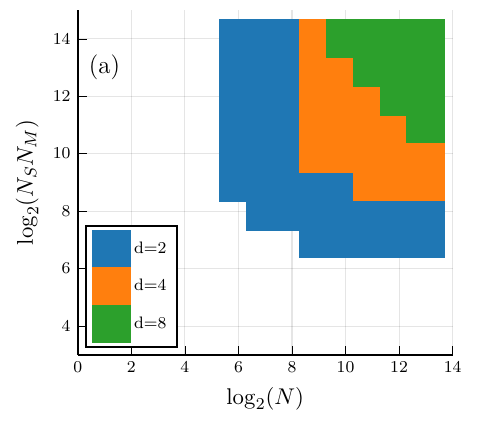}
	\includegraphics[width=0.45\textwidth]{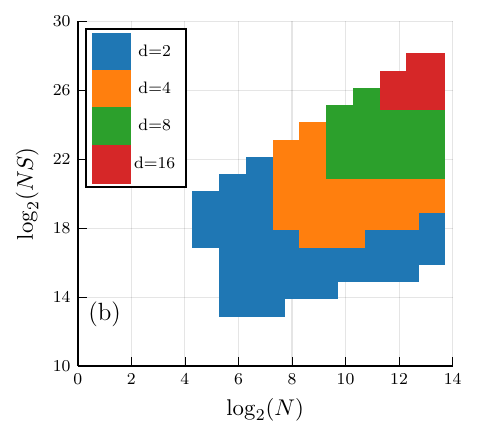}
	\caption{
		(a) Performance of the learning approach, as in Fig.~\ref{fig:success}(a), but 
		with the swap test replaced by the shadow overlap estimator with $N_U$ unitaries 
		and $N_M$.
		For $d=16$ (not shown) the maximum success rate is 96\%, which is below the threshold of 
		99\%. 
		(b) Same data of Fig.~\ref{fig:success}(a), but reshaped to have $\log_2(NS)$ in 
		the vertical axis. 
	}
	\label{fig:shadow}
\end{figure}

\section{Further numerical experiments} \label{sec:more}

Numerical simulations with support vector machines are shown in Fig.~\ref{fig:result}. 
We performed numerical experiments for different dimensions $d$ 
of the Hilbert spaces of the two subsystems in Eq.~\eqref{eq:states}.
From the numerical simulations, we see that for $d=2$ and $d=4$, the success rate is low
for low values of $S$ and $N$, but then reaches basically perfect success 
for larger values of both $N$ and $S$. 
On the other hand, for the still small value $d=32$, namely when the subsystems in \eqref{eq:states} 
have just 5 qubits, the success rate is close to that of random decisions (50\%), in spite of 
the large number of training data (up to $\sim8000$) and large number of shots ($\sim16000$). 
When $N$ and $S$ are both large, we see that the accuracy starts to increase up to $\sim 70\%$. 

The performance of the shadow estimator is then shown in Fig.~\ref{fig:shadow}(a), while that 
of the swap test is shown in Fig.~\ref{fig:shadow}(b). In those figures we chose we put in the 
vertical axis the number of copies of each state that are required in the two methods. 
For the swap test this is $NS$, since during during, the overlap with all training states 
must be estimated, while for the shadow estimator this is $N_MN_U$, irrespective of the number 
of training states.

\section{The twirling superoperator}
Given the unitary group acting in the Hilbert space 
$\mathcal H$,  the 
twirling superoperator is defined by
\begin{equation}
	\mathcal T^n(A) = \int dU\; {U^{\otimes n}}A\,
	(U^\dagger)^{ \otimes n},
  \label{e.twirling}
\end{equation}
where $A$ is an operator in $\mathcal H^{\otimes n}$ and $dU$ is the Haar
measure over the unitary group, which satisfies $dU = dUV =d VU$ for each
unitary matrix $V$, and is normalized such that $\int dU = \openone$.
The twirling superoperator is an important tool in quantum information that 
appears in many proofs. From the invariance of the Haar measure under unitary transformation, one finds that 
the twirling operator commutes with each $V^{\otimes n}$ 
\begin{align}
	V^{\otimes n}\,\mathcal T^n(A) \,  V^{\otimes n\dagger} = 
	\mathcal T^n(A),
\end{align}
and, accordingly, from the Weyl-Schur duality \cite{boerner1970representations} 
it is a linear combination of index permuting 
operators $  \mathcal T^n(A) = \sum_\pi b_\pi\; P_\pi$, where 
each $\pi$ belongs to the group $S_n$ of permutations.
Multiplying the two sides by $P_\sigma$, and taking the trace 
we get
\begin{equation*}
  a_\sigma = \Tr \left[P_\sigma\,\mathcal T^n(A)\right] = \Tr\left[
  P_\sigma\,A\right] = \sum_\pi b_\pi \; \Tr\left[P_\sigma\,P_\pi\right],
\end{equation*}
which can be expressed as a linear system of equations $\sum_\pi M_{\sigma\pi} b_\pi = a_\sigma$,
with the matrix $M_{\sigma\pi}=\Tr\left[P_\sigma\,P_\pi\right]$.
The second equality comes from the definition in Eq.~(\ref{e.twirling}) and the fact the index permuting operators commute with unitaries of the form $U^{\otimes n}$.
Even if the
matrix $M$ is singular we can define $M^{-1}$ as the Moore-Penrose 
\emph{pseudo-inverse},
which is nothing but the standard inverse when $M$ is invertible, 
and accordingly 
\begin{equation}
   \mathcal T^n(A) = 
  \sum_{\pi,\sigma} \left(M^{-1}\right)_{\pi\sigma} \; 
  \Tr\left[ P_\sigma\,A\right] \; P_\pi.
  \label{e.twirlingresult}
\end{equation}
A closed-form expression for the $M$ matrix can be obtained using tools 
from representation theory 
\cite{brandao2011convergence} 
\begin{equation}
	M_{\sigma\pi} = \Tr[P_\sigma P_\pi] = d^{\odot(\sigma\pi^{-1})},
	\label{eq:M mat}
\end{equation}
where $\odot(\sigma)$ is the number of cycles in the cycle decomposition
of $\sigma \in S_n$. For low values of $n$, it can also be computed 
explicitly. For instance: 

{\it Case $n=1$}: there is a single permutation operator $P_0$, 
which is the identity. Hence $\Tr[P_0]=d$ and 
\begin{equation}
	\mathcal T^1(A) = \frac{\Tr[A]}d \openone.
	\label{eq:twirln1}
\end{equation}

{\it Case $n=2$}:
there are only two index permuting operators: the identity 
operator $P_0 =\openone$ and the swap operator $P_1=S$.
Since $\Tr P_0 = d^2$ and $\Tr P_1 = d$ we directly get
\begin{align*}
  M=\begin{pmatrix}
    d^2 & d \\ d & d^2 
  \end{pmatrix}, 
  && \Longrightarrow &&
  M^{-1}=\frac1{d(d^2-1)}\begin{pmatrix}
    d & -1 \\ -1 & d
  \end{pmatrix}, 
\end{align*}
and thus
\begin{equation}
  \mathcal T^2(A) = \frac{d\Tr A - \Tr\left[SA\right]}{d(d^2-1)}\; 
  \openone
	+ \frac{d\Tr\left[SA\right]-\Tr A}{d(d^2-1)}\; S.
  \label{e.twirln2}
\end{equation}

{\it Asymptotics:} For large $d$, it holds that \cite{collins2006integration}
\begin{equation}
(M^{-1})_{\sigma\pi} = \mathcal O\left(d^{-n-\lfloor\sigma^{-1}\pi\rfloor}\right),
\label{eq:twirl asymp}
\end{equation}
where $\lfloor\sigma\rfloor$ denotes the minimal number of factors necessary to
write $\sigma$ as a product of
transpositions. Therefore, 
$(M^{-1})_{\sigma\pi}$ is approximately diagonal for large $d$, with $\mathcal O(d^{-n})$ diagonal entries.

\section{Average states} \label{sec:c2}
Using the twirling superoperator the average separable and 
entangled states can be explicitly computed as 
\begin{equation}
\begin{aligned}
	\bar \rho^{\rm sep} &= \mathbb E[\rho^{\rm sep}] = (\mathcal T_A^1\otimes \mathcal T_B^1)
	[\ket{00}\!\bra{00}] = \frac{\openone}{d^2}, \\ 
	\bar \rho^{\rm ent} &= \mathbb E[\rho^{\rm ent}] = (\mathcal T_A^1\otimes \mathcal T_B^1)
	[\ket{\Phi}\!\bra{\Phi}] = \frac{\openone}{d^2},
\end{aligned}
\label{eq:single copy}
\end{equation}
showing that the average states are equal and hence not distinguishable. 
However, they become distinguishable once we consider multiple copies.
For instance,
\begin{equation}
	\bar \rho^{\rm sep}_{(2)} = \mathbb E[(\rho^{\rm sep})^{\otimes 2}] =
	\mathcal T^2[\ket{00}\!\bra{00}]^{\otimes 2} = \frac{(\openone +S)^{\otimes 2}}{d^2(d+1)^2}.
	\label{eq:ave rho sep 2}
\end{equation}
We must be careful about the order of the systems, since this tells us which systems we twirl together and which we twirl independently: here they take the order $A_1 A_2 B_1 B_2$.
The computation of the average entangled state is a bit more tedious, though straightforward 
\begin{align}
	\bar \rho^{\rm ent}_{(2)} 
	&= \mathbb E[(\rho^{\rm ent})^{\otimes 2}] = 
	(\mathcal T_A^2\otimes \mathcal T_B^2)[\Phi_{A_1,B_1}\Phi_{A_2,B_2}]\label{eq:ave rho ent 2}
	\\ & = \nonumber
	\frac1{d^2} \sum_{ijkl} \mathcal T^2[\ket{ij}\!\bra{kl}]^{\otimes 2}
	\\&= \nonumber
	\sum_{ijkl} \frac{[(d \delta_{ik}\delta_{jl}-\delta_{il}\delta_{jk}) \openone + 
	(d \delta_{il}\delta_{jk}-\delta_{ik}\delta_{jl}) S]^{\otimes 2}}{d^4(d^2-1)^2} 
	\\&= \nonumber
	\frac{(d^2d^2{+}d^2{-}2d^2)(\openone{+}S^{\otimes 2})+(d^2d{+}d{-}2dd^2)(S_A{+}S_B)}{d^4(d^2-1)^2} 
	\\&= \nonumber
	\frac{\openone + S^{\otimes 2}}{d^2(d^2-1)} -  \frac{S_A + S_B }{d^3(d^2-1)},
\end{align}
where $\Phi = \ket\Phi\!\bra\Phi$, $A_{1,2}$ and $B_{1,2}$ respectively refer to the two copies 
of $A$ or $B$, and $S_A$ and $S_B$ are
$S\otimes \openone$ and $\openone\otimes S$ respectively.
Although in the first line we use the ordering $A_1 B_1 A_2 B_2$ for the term inside the square brackets, in the remaining lines we always use the system ordering $A_1 A_2 B_1 B_2$.
In the fourth line we use the simple identities $\sum_{ijkl} \delta_{ik}\delta_{jk}=d^2$ 
and $\sum_{ijkl}\delta_{ik}\delta_{jl}\delta_{il}\delta_{jk} = d$.

To simplify the comparison of the two average states, it is useful to introduce the 
projections $P_\pm = (\openone \pm S)/2$ onto the symmetric and antisymmetric 
subspaces. By explicit calculation, those subspaces have dimension 
\begin{align}
	d_+ &= \Tr[P_+] = \frac{d(d+1)}2, &
	d_- &= \Tr[P_-] = \frac{d(d-1)}2.
\end{align}
Then (using system order $A_1 A_2 B_1 B_2$)
\begin{align}
	\bar \rho^{\rm sep}_{(2)}  &= \frac{P_+\otimes P_+}{d_+^2},\label{eq:rhoSep} \\
	\bar \rho^{\rm ent}_{(2)}  &= \frac1{d^2}\left(\frac{P_+\otimes P_+}{d_+}+\frac{P_-\otimes P_-}{d_-}\right).\label{eq:rhoEnt}
\end{align}
From the above expressions we can compute the purities as 
\begin{align}
	\Delta_{_{++}} &:=
	\Tr[	(\bar \rho^{\rm sep}_{(2)})^2] = \frac{d_+^2}{d_+^4} = \frac{4}{d^2(d+1)^2},\label{eq: overlap ++} \\
	\Delta_{_{--}}&:=\Tr[	(\bar \rho^{\rm ent}_{(2)})^2] = \frac{1}{d^4}\left(\frac{d_+^2}{d_+^2}+\frac{d_-^2}{d_-^2}\right) = \frac{2}{d^4},\label{eq: overlap --}
\end{align}
as well as the overlap 
\begin{equation}
 \Delta_{_{+-}} := 
	\Tr[\bar \rho^{\rm sep}_{(2)}\bar \rho^{\rm ent}_{(2)}]  = \frac{d_+^2}{d_+^3d^2}=\frac{2}{d^3(d+1)}.\label{eq: overlap +-}
\end{equation}

From
Eqs.~(\ref{eq:rhoSep}) and (\ref{eq:rhoEnt}),
we can compute the average state 
\begin{equation}
	\bar \rho_2 = \frac{	\bar \rho^{\rm sep}_2  + \bar \rho^{\rm ent}_2}2 =  
	\frac{1}{d^3}
 \left(
 \frac{3d+1}{(d+1)^2}P_+\otimes P_+
 +\frac{P_-\otimes P_-}{d-1}
 \right).
\end{equation}
and 
\begin{align}
	\Tr[\sqrt{\bar\rho_2}] &= \frac{d_-^2}{\sqrt{d^3(d-1)}} + \frac{\sqrt{3d+1}}{d+1}\frac{d_+^2}{\sqrt{d^3}}
	\nonumber \\ & \simeq \frac{d^4}{4d^2}+ \frac{\sqrt{3}d^4}{4 d^2} = \frac{\sqrt 3+1}{4}d^2. 
\end{align}
From the above result we can compute the upper bound
$\sqrt{\mathcal B/N}$ on the 
generalization error \cite{banchi2021generalization,banchi2024statistical}, 
which implies that, in order 
in order to ensure a small generalization error, we need a number of samples $N$ 
that satisfies 
\begin{equation}
	N \gg \mathcal B = (\Tr[\sqrt{\bar\rho_2}])^2 = \mathcal O(d^4). 
	\label{eq:gen bound}
\end{equation}

On the other hand, within  the single-copy subspace, using Eqs.~\eqref{eq:single copy}, we find 
$\mathcal B = (\Tr[(\rho^{\mathrm{sep}}+\rho^{\mathrm{ent}})/2] )^2 = d^2$.

	We can also use Eqs.~(\ref{eq:rhoSep}) and (\ref{eq:rhoEnt}) to calculate the trace norm between the average states and so show that the SWAP test is the optimal measurement for discriminating between $\rho^{\rm sep}_{(2)}$ and $\rho^{\rm ent}_{(2)}$.

We start by calculating
\begin{align}
	\nonumber
        \left\|\bar \rho^{\rm sep}_{(2)} - \bar \rho^{\rm ent}_{(2)}\right\|
        &= \frac{d^2-d_+}{d^2 d_+^2}\left\|P_+\otimes P_+\right\|+\frac{1}{d^2 d_-}\left\| P_-\otimes P_- \right\| \\
        &= \frac{d^2-d_+}{d^2}+\frac{d_-}{d^2}
        =1-\frac{1}{d},
\end{align}
where, on the first line, we use the fact that $P_+$ and $P_-$ exist in orthogonal subspaces. Then, the maximum average success probability, for a single measurement, is
\begin{equation}
    p_{\mathrm{succ}}=\frac{1}{2}-\frac{1}{4}\left\|\bar \rho^{\rm sep}_{(2)} - \bar \rho^{\rm ent}_{(2)}\right\|
    = \frac{3}{4} - \frac{1}{4d}.
\end{equation}

The SWAP measurement is the POVM with operators $\{P_+,P_-\}$, with $P_\pm=(\openone\pm {\rm SWAP})/2$. 
In our case, we carry out the SWAP measurement on the $A$ modes, whilst tracing over the $B$ modes. The measurement is successful if we attain a result of $0$ for $(\rho^{\rm sep})^{\otimes 2}$ or $1$ for $(\rho^{\rm ent})^{\otimes 2}$. These outcomes have average probabilities
\begin{align}
    &\Tr[(P_+\otimes\openone ) \rho^{\rm sep}_{(2)}] = 1,\\
    &\Tr[(P_-\otimes\openone ) \rho^{\rm ent}_{(2)}] = \frac{d_+}{d^2} = \frac{d+1}{2d},
\end{align}
and since each class has equal probability, the overall success probability is $\frac{3}{4} - \frac{1}{4d}$. Hence, this is the optimal measurement in terms of success probability.

Finally, by using Eqs.~(\ref{eq:ave rho sep 2}) and (\ref{eq:ave rho ent 2}), we can show that an observable of the form in Eq.~(\ref{eq:decision obs emp}) can approach Eq.~(\ref{eq:representer}) in the limit of large $N$. We do this by replacing the empirical averages in Eq.~(\ref{eq:decision obs emp}) with the true average states that we have calculated. By choosing the parameters $\alpha_{\pm}$ and $\beta$ correctly, we can recover the observable in Eq.~(\ref{eq:representer}), thus demonstrating that it is possible to reach this observable via the representer theorem. A set of parameter choices that shows this is
\begin{equation}
    \alpha_+ = (d+1),
    \quad \alpha_- = (d-1),
    \quad \beta = \frac{d+1}{d}.
\end{equation}

\section{Error analysis with average states}\label{sec:error}

Following Ref.~\cite{banchi2024statistical},
by exploiting the convexity of the hinge loss $\ell$ we may bound the abstract loss as 
\begin{align}
	L &= \sum_y \int d\mu(\rho^y)\; \ell(y,f[(\rho^y)^{\otimes c})]) + \|A\|_2^2
 \nonumber \\ &
 \leq \sum_y  \ell\left(y,f(\bar\rho^y_{(c)})\right) + \|A\|_2^2,
 \label{eq:Loss ineq}
\end{align}
where $d\mu(\rho)$ represents the measure over the states \eqref{eq:states} with Haar-uniform 
unitaries $U_{A/B}$, and 
\begin{equation}
	\bar \rho^y_{(c)} = \int d\mu(\rho^y)\; (\rho^y)^{\otimes c},
	\label{eq:ave states}
\end{equation}
are the average states. 
The inequality \eqref{eq:Loss ineq} is like \eqref{eq:Loss emp ineq}, but with formal averages over 
the full distribution, rather than the empirical averages over the $2N$ training states. 
From the inequality \eqref{eq:Loss ineq}, we can bound the average loss over the individual states 
in terms of the loss over the average states \eqref{eq:ave states}. 
By inserting a $\|A-b\openone\|_2^2$ penalty term, 
the representer theorem guarantees that the optimal observable is like 
Eq.~\eqref{eq:decision obs emp}, 
	$\bar A^*= \alpha_+\bar \rho^+_{(c)}+\alpha_-\bar\rho^-_{(c)} +b$,
	where now the optimal coefficients can be computed analytically \cite{banchi2024statistical}. We find 
\begin{equation}
	\bar A^*_{(c)}= \frac{2\left(\bar \rho^+_{(c)}-\bar\rho^-_{(c)}\right)-\left(\Delta_{_{++}}-\Delta_{_{--}}\right)\openone}{\Delta_{_{++}}+\Delta_{_{--}}-2\Delta_{_{-+}}},
	\label{eq:opt A bar}
\end{equation}
where $\Delta_{yy'} = \Tr[	\bar \rho^y_{(c)}\bar \rho^{y'}_{(c)} ]$ are the average kernel entries
\begin{equation}
\begin{aligned}
	\Delta_{_{++}} &= \frac{4}{d^2(d+1)^2}, 
								 &
	\Delta_{_{--}} &= \frac{2}{d^4},
								 \\
	\Delta_{_{+-}} &= \Delta_{_{-+}} 
								 = \frac{2}{d^3(d+1)}.
\end{aligned}
\label{eq:average kerneles}
\end{equation}
which were obtained in Appendix~\ref{sec:c2}. 
Exploiting similar analytical expressions (see Appendix~\ref{sec:c2}),
we get for $c=2$
\begin{equation*}
	\bar A^*_{(2)} = 
    \frac{d}{d-1}\bigg(S_A+S_B - \frac{2d S_{AB}}{d^2 +1} 
    + \Big(1-\frac{d^2 - 1}{d(d^2+1)}\Big)\openone \bigg)
\end{equation*}
where $S_A={\rm SWAP}_{A_1,A_2}$ with the identity on the $B$ modes, vice versa
for $S_B$, and $S_{AB}$ is the SWAP operator applied independently to both
systems.
Without loss of generality, we may replace $S_{AB}$ with an identity operator, since 
the expectation value of the $S_{AB}$ over pure states is always 1. In this way, 
$\bar A^*_{(2)}$ becomes  identical to the optimal observable \eqref{eq:representer}. 
This shows that the latter can be obtained by minimizing the loss over 
average states \eqref{eq:Loss ineq}, thus completing the proof of Theorem~\ref{thm:optimal}. 

The upper bound from \eqref{eq:Loss ineq} allows us to focus on 
the average states, which are simpler to study than the individual states \eqref{eq:states}. 
In the learning scenario where we have $N$ different samples, we may apply the same 
inequality \eqref{eq:Loss ineq} to the empirical loss \eqref{eq:L emp}. The optimal observable
$\hat A^{*}_{(c)}$,
according to the resulting bound, is then the one obtained by replacing 
the abstract averages \eqref{eq:ave states} in all the terms in \eqref{eq:opt A bar} with 
the empirical averages \eqref{eq:emp state}, which are distinguished by a ``hat'' symbol. 
At test stage, we need to test the performance of the empirical observable 
$\hat A^{*}_{(c)}$ with new data, which are distributed as in \eqref{eq:Loss ineq}.
Therefore, we may bound the test loss, for $c=2$, as 
$	\sum_y\ell\left(y,\Tr\big[\bar \rho_{(2)}^y \hat A^{*}_{(2)}\big]\right)$.
Wrong classification of a test state $\rho^y$ happens when 
\begin{equation}
	{\rm sign}\left(\Tr\big[\rho^{y\otimes 2} \hat A^{*}_{(2)}\big]\right) 
\neq y.
\label{eq:sign prediction}
\end{equation}
Since the sign is independent of the overall denominator,
we may alternatively study $\Tr\big[\bar \rho^{y\otimes 2} \hat B^{}_{(2)}\big]$, where 
$\hat B^{}_{(2)}$ has been defined in \eqref{eq:opt B}.

\subsection{Training and testing: two-copy case} 

We now focus on studying the reliability of the operator in Eq.~\eqref{eq:opt B}.
For any test state $\rho^y$, we can calculate the expected value of 
our estimator of $B^{}_{(2)}$, $B_{\mathrm{obs}}$ as
\begin{equation}
	\Tr[\rho^{y\otimes 2} B_{\mathrm{obs}}] = 2\left(\hathat\Delta_{_+}^y- \hathat\Delta_{_-}^y\right) - 
	\left(\hathat \Delta^{}_{_{++}}-\hathat \Delta^{}_{_{--}}\right),
\end{equation}
where $\hathat \Delta$ are ``doubly-empirical'' estimations via swap tests with $S$ shots and $N$ training states
and $\Delta_{_\pm}^y$ refers to the overlap of $\rho^{y\otimes 2}$ with the average states $\rho^{\rm sep}_{(2)}$ and $\bar \rho^{\rm ent}_{(2)}$ (or their empirical approximations).

\begin{thm}\label{thm:two copies}
	\begin{equation}
		\hathat \Delta_{\alpha\alpha} = \frac{2}{S N(N-1)}\sum_{\substack{i,j=1 \\ i<j}}^N \sum_{s=1}^S (-1)^{o_{ij\alpha s}}
		\label{eq:Delta hathat training}
	\end{equation}
is an unbiased estimator of $\Delta_{\alpha\alpha}$, 
where $o_{ij\alpha s}$ are the measurement outcomes of $S$ swap tests, each performed 
over two copies of the training states $\rho^{\alpha\otimes2}_i$ and $\rho^{\alpha\otimes 2}_j$. 
Similarly,
	\begin{equation}
		\hathat \Delta^y_{\alpha} = \frac{1}{S N}\sum_{i=1}^N \sum_{s=1}^S (-1)^{o^y_{i\alpha s}}
		\label{eq:Delta hathat testing}
	\end{equation}
	is an unbiased estimator of $\Tr[\rho^{y\otimes 2}\bar \rho_{(2)}^\alpha]$, 
where $o^y_{i\alpha s}$ are the measurement outcomes of $S$ swap tests, each performed 
over two copies of the training states $\rho^{\alpha\otimes2}_i$ and test state $\rho^{y\otimes 2}$. 
\end{thm}
\noindent{\it Proof.} By definition 
\begin{equation}
	\mathbb E_{\mathrm{shots}} [\hathat\Delta_{\alpha\alpha}] =\frac{2}{N(N-1)}\sum_{i<j} \Tr[\rho_i^{\alpha\otimes 2}\rho_j^{\alpha\otimes 2}]. 
\end{equation}
Since different training states are independent and identically distributed, by taking an average over them we get
\begin{align}
	\nonumber \mathbb E_{\mathrm{train, shots}} [\hathat\Delta_{\alpha\alpha}]  
	&= \frac{2}{N(N-1)}\sum_{i<j} \Tr[\bar\rho^{\alpha}_{(2)}\bar\rho^{\alpha}_{(2)}]
\\&= \Tr[\bar\rho^{\alpha}_{(2)}{}^2] = \Delta_{\alpha\alpha}.
\end{align}
Similarly $\mathbb E_{\mathrm{train, shots}} [\hathat\Delta_{\alpha}^y] =
\Tr[\rho^{y\otimes 2}\bar \rho_{(2)}^\alpha]$. 
\hfill$\qed$

\vspace{5mm}
To compute the variance of the above estimators we evaluate
\begin{align}
	\mathbb E[(\hathat\Delta_{\alpha\alpha})^2] 
	&=  C^{-2} \sum_{i<j,i'<j'}\sum_{s,s'} \mathbb E[(-1)^{o_{ij\alpha s}+o_{i'j'\alpha s'}}] \nonumber
	\\  &=
		\Delta_{\rm same}^2 + \Delta^2_{\rm diff},
	\label{eq:Delta emp same diff}
\end{align}
where $C=N(N-1)S/2$, 
$\mathbb E \equiv \mathbb E_{\mathrm{train, shots}} $  refers to the expected value taken over both
the measurement outcomes and the choice of training set and where
we split the sum among the contributions from the ``same experiment''
\begin{align}
	\Delta_{\rm same}^2 &= 
	C^{-2}\mathbb E\left[\sum_{i<j} \sum_s(1) + \sum_{s\neq s'} \Tr[\rho_i\rho_j]^4 \right]   \\
											&= 
	\frac{N(N-1) S/2 + S(S-1)\sum_{i<j}\mathbb E[\Tr[\rho_i\rho_j]^4] }{C^2},
	\nonumber
\end{align}
where we dropped the dependence on $\alpha$ to simplify the notation,
 and the contributions 
\begin{align}
	\Delta^2_{\rm diff} &= \nonumber
	C^{-2}\mathbb E\left[\sum_{\substack{i<j,i'<j'\\(ij)\neq(i'j')}}\sum_{s,s'} 
	\Tr[\rho_i\rho_j]^2 \Tr[\rho_{i'}\rho_{j'}]^2\right] 
									 \\ &= 
									 \frac{S^2}{C^2}\mathbb E_{\rm set}\left[\sum_{\substack{i<j,i'<j'\\(ij)\neq(i'j')}}
	\Tr[\rho_i\rho_j]^2 \Tr[\rho_{i'}\rho_{j'}]^2 \right]
\end{align}
due to independent experiments performed with different pairs of states. 
The notation $(ab)\neq(cd)$ means that the pair is different, yet some numbers 
might be equal, e.g.,~$a$ could be equal to $c$ or $d$, provided that the pair is different. 

Regarding the first term, we find 
\begin{align}
	\Delta_{\rm same}^2 &= \frac{1 +(S-1) \mathbb E_{\rm set}
	\left[\Tr[\rho\rho']^4\right]}C.
\end{align}
Since the individual states are pure, we can write
\begin{align}
    \mathbb E\left[\Tr[\rho\rho']^m\right]
    = \Tr\left[\left(\mathcal{T}^m [\sigma^{\otimes m}]\right)^2\right]
    \lesssim \mathcal O(d^{-2m}),\label{eq: twirling cont}
\end{align}
where $\sigma$ could be $\ket{00}\!\bra{00}$ or $\ket{\Phi}\!\bra{\Phi}$ and where the inequality uses
Eq.~\eqref{eq:twirl asymp}, to show that the twirling
introduces $d$-dependent corrections.
Hence, the dominant term is 
\begin{equation}
      \Delta_{\rm same}^2\approx \frac1C \approx\frac2{N^2S}. 
\end{equation}
More precisely, from generalized mean inequalities, 
$ \mathbb E_{\rm set} \left[\Tr[\rho\rho']^2\right]^2  \leq 
 \mathbb E_{\rm set} \left[\Tr[\rho\rho']^4\right] \leq
 \mathbb E_{\rm set} \left[\Tr[\rho\rho']^2\right]$, so 
\begin{equation}
	\frac2{N^2S} + \frac{2}{N^2} \Delta_{\alpha\alpha}^2 \lesssim
      \Delta_{\rm same}^2\lesssim
	\frac2{N^2S} + \frac{2}{N^2} \Delta_{\alpha\alpha},
	\label{eq:Delta same ineq}
\end{equation}
which shows that, even for $S\to\infty$, the variance is still non-zero.

Similarly, by using Eq.~\eqref{eq:twirl asymp}, we see that $\Delta^2_{\rm diff}$ only contains 
$d$-dependent corrections.
Therefore, 
\begin{equation}
	\mathbb E\left[ 
	{\rm Var}[ \hathat\Delta_{\alpha\alpha}]\right] \simeq \mathcal O\left[\frac1{N^2 S }
\right] 
	,\label{eq:Delta 2 copy} 
\end{equation}
where the first term can 
be intuitively understood since there are $\mathcal O(N(N-1) S/2)$ independent swap tests, 
each with unit variance. 

Similarly,
\begin{align}
	\mathbb E[ (\hathat\Delta^y_{\alpha})^2] 
	&= \frac1{N^2S^2} \sum_{i,s}\sum_{i's'} \mathbb E[(-1)^{o_{is\alpha}^y+o_{i's'\alpha}^y}]
	\\ \nonumber
	&= \frac1{N^2S^2}\mathbb E\left[\sum_{i} \sum_s(1) + \sum_{s\neq s'} \Tr[\rho^y\rho_i^\alpha]^4 \right]  
	\\\nonumber 
	&+ \frac1{N^2S^2}\mathbb E\left[\sum_{i\neq i'}\sum_{s,s'} 
	\Tr[\rho_i^\alpha\rho^y]^2 \Tr[\rho_{i'}^\alpha\rho^y]^2\right],
\end{align}
from which we get 
\begin{equation}
	{\rm Var}[ \hathat\Delta^y_{\alpha}]  \simeq 
	\mathcal O\left[\frac1{NS}
	\right] .
	\label{eq: Delta test 2}
\end{equation}

\section{Swap test with single-copy states}\label{sec:single copy}
We now consider different estimators that do not make use of two copies 
of each state, but rather estimate $\Tr[\rho\sigma]^2$ by first 
estimating $\Tr[\rho\sigma]$ using a single copy of $\rho$ and $\sigma$, and then taking the square. 
\begin{thm}\label{thm:single}
	\begin{align}
		\hathat \Delta_{\alpha\alpha} &= 
		\frac{2S}{(S-1) N(N-1)}\sum_{\substack{i,j=1 \\ i<j}}^N \left[\left(\sum_{s=1}^S \frac{(-1)^{o_{ij\alpha s}}}S\right)^2-\frac{1}{S^2}\right]
\nonumber
		\\&=
		\frac{2}{N(N-1)}\sum_{\substack{i,j=1 \\ i<j}}^N \sum_{\substack{s,t=1\\s\neq t}}^S \frac{(-1)^{o_{ij\alpha s}+o_{ij\alpha t}}}{S(S-1)}
		\label{eq:Delta hathat training 1}
	\end{align}
is an unbiased estimator of $\Delta_{\alpha\alpha}$, 
where $o_{ij\alpha s}$ are the measurement outcomes of $S$ swap tests 
over the training states $\rho^{\alpha}_i$ and $\rho^{\alpha}_j$. 
Similarly,
	\begin{align}
		\hathat \Delta^y_{\alpha} &= \nonumber
		\frac{S}{(S-1)N}\sum_{i=1}^N \left[\left(\sum_{s=1}^S \frac{ (-1)^{o^y_{i\alpha s}}}S\right)^2-\frac1{S^2}\right]
														\\&=
		\frac{1}{S (S-1)N}\sum_{i=1}^N \sum_{\substack{s,t=1\\s\neq t}}^S  (-1)^{o^y_{i\alpha s}+o^y_{i\alpha t}}
		\label{eq:Delta hathat testing 1}
	\end{align}
	is an unbiased estimator of $\Tr[\rho^{y\otimes 2}\bar \rho_{(2)}^\alpha]$, 
where $o^y_{i\alpha s}$ are the measurement outcomes of $S$ swap tests 
over the training states $\rho^{\alpha}_i$ and test state $\rho^{y}$. 
\end{thm}
\noindent{\it Proof.} Since the shots with $s\neq t$ are independent and identically distributed 
\begin{align}
	\mathbb E_{\mathrm{shots}} [\hathat\Delta_{\alpha\alpha}] &=\frac{2}{N(N-1)}\sum_{i<j} 
	\Tr[\rho_i^{\alpha}\rho_j^{\alpha}]^2 \nonumber \\ &=
\frac{2}{N(N-1)}\sum_{i<j} 
	\Tr[\rho_i^{\alpha\otimes 2}\rho_j^{\alpha\otimes 2}]. 
\end{align}
Then the proof follows that of Theorem~\ref{thm:two copies}. 
\hfill$\qed$

\vspace{5mm}
To compute the variance we evaluate
\begin{align}
	\mathbb{E}&[\left(\hathat\Delta_{\alpha\alpha}\right) ^2] = \\&   \nonumber
	G^{-2} \sum_{i< j, i'< j'}^N \sum_{s\neq t,s'\neq t'}^S  
	\mathbb{E}[(-1)^{o_{ijs}+o_{ijt}+o_{i'j's'}+o_{i'j't'}}],
\end{align}
where $G={S(S-1) N(N-1)}/2$. 
We can now split the above sum, as in Eq.~\eqref{eq:Delta emp same diff}, as the 
contribution coming from the ``same experiment'' with $i=i'$ and $j=j'$, 
and ``different experiments'' when 
$(ij)\neq(i'j')$. 
\begin{align}
	\Delta_{\mathrm{same}}^2& = \nonumber
	G^{-2} \sum_{i<j}^N \sum_{s\neq t,s'\neq t'}^S  
	\mathbb{E}[(-1)^{o_{ijs}+o_{ijt}+o_{ijs'}+o_{ijt'}}]
	\\& \nonumber
	\simeq \frac{2+4S \Delta_{\alpha\alpha}}G + \frac{S^4N(N-1)}{2G^2} \mathbb E_{\rm set} \left[\Tr[\rho\rho']^4\right],
	\\& 
	\simeq \frac{2+4S \Delta_{\alpha\alpha} + S^2 \mathbb E_{\rm set} \left[\Tr[\rho\rho']^4\right]}G,
\end{align}
where the first, dominant term comes from when $s=s'$ and $t=t'$ or $s=t'$ and $t=s'$,
the second one comes when $s=t'$, or $s=t$, or $s'=t'$, or $s'=t$, and the other two indices 
are different, while the third term comes from when all indices are different. 
The approximation comes in displaying only the dominant term in $S$.
With a similar analysis we find that the contribution coming from ``different experiments'' 
introduces $d$-dependent corrections. Therefore, 
\begin{equation}
	\mathrm{Var}[\hathat\Delta_{\alpha\alpha}] 
	\simeq \mathcal O\left[\frac1{N^2}\left(\frac{1}{S^2} + \frac1{S d^4} + \frac1{d^8}\right)\right]
			\label{eq:Delta 1 copy}
\end{equation}
which, for large $S$,  is smaller than Eq.~\eqref{eq:Delta 2 copy}.

As for the variance at test stage, 
\begin{align}
	\mathbb E[& (	\hathat \Delta^y_{\alpha} )^2]  \\ \nonumber
						&\simeq \frac1{N^2S^4} \sum_{ii'}\sum_{s\neq t}\sum_{s'\neq t'} 
						\mathbb E[(-1)^{o^y_{is\alpha}+o^y_{i's'\alpha}+o^y_{it\alpha}+o^y_{i't'\alpha}}],
\end{align}
and proceeding as in the previous case we get 
\begin{equation}
	\mathrm{Var}[\hathat \Delta^y_{\alpha} ]   
	\simeq \mathcal O\left[\frac1{N}\left(\frac{1}{S^2} + \frac1{S d^4} + \frac1{d^8}\right)\right],
	\label{eq: Delta test 1 a}
\end{equation}
which, for large $S$,  is smaller than Eq.~\eqref{eq: Delta test 2}.

\end{document}